\documentclass[useAMS,usenatbib]{mnras}

\usepackage[]{color,graphicx}

\usepackage{amsmath}

\def\lesssim{\mathrel{\hbox{\rlap{\hbox{\lower5pt\hbox{$\sim$}}}\hbox{$<$}}}}
\def\gtrsim{\mathrel{\hbox{\rlap{\hbox{\lower5pt\hbox{$\sim$}}}\hbox{$>$}}}}

\title[Neutrino-driven explosions of ultra-stripped type Ic
  supernovae]{Neutrino-driven explosions of ultra-stripped type Ic
  supernovae generating binary neutron stars}

\author[Suwa et al.]{
Yudai~Suwa$^{1,2}$\thanks{E-mail: suwa@yukawa.kyoto-u.ac.jp},
Takashi~Yoshida$^{1}$,
Masaru~Shibata$^{1}$,
Hideyuki~Umeda$^{3}$,
\newauthor
and Koh~Takahashi$^{3}$
\\
$^{1}$Yukawa Institute for Theoretical Physics, Kyoto University,
Oiwake-cho, Kitashirakawa, Sakyo-ku, Kyoto, 606-8502, Japan\\
$^{2}$Max-Planck-Institut f\"ur Astrophysik,
  Karl-Schwarzschild-Str. 1, D-85748 Garching, Germany\\
$^{3}$Department of Astronomy, Graduate School of Science, University
  of Tokyo, Tokyo 113-0033, Japan\\
}

\begin{document}

\date{Accepted. Received.}

\pagerange{\pageref{firstpage}--\pageref{lastpage}} \pubyear{2015}

\maketitle

\label{firstpage}

\begin{abstract}
We study explosion characteristics of ultra-stripped supernovae (SNe),
which are candidates of SNe generating binary neutron stars (NSs). As
a first step, we perform stellar evolutionary simulations of bare
carbon-oxygen cores of mass from 1.45 to 2.0 $M_\odot$ until the iron
cores become unstable and start collapsing. We then perform
axisymmetric hydrodynamics simulations with spectral neutrino
transport using these stellar evolution outcomes as initial
conditions. All models exhibit successful explosions driven by
neutrino heating. The diagnostic explosion energy, ejecta mass, Ni
mass, and NS mass are typically $\sim 10^{50}$ erg, $\sim 0.1
M_\odot$, $\sim 0.01M_\odot$, and $\approx 1.3 M_\odot$, which are
compatible with observations of rapidly-evolving and luminous
transient such as SN 2005ek. We also find that the ultra-stripped SN
is a candidate for producing the secondary low-mass NS in the observed
compact binary NSs like PSR J0737-3039.
\end{abstract}


\section{Introduction}
\label{sec:intro}

Mergers of binary compact objects, i.e. neutron stars (NSs) and black
holes (BHs), are promising candidates of strong gravitational wave
(GW) sources. Event rates of these mergers are estimated based on the
number of observed binary NSs in our galaxy and population synthesis
calculations \citep[e.g.,][]{abad10}. These estimates, however, have
large uncertainty with, roughly speaking, two orders of magnitude.
Recalling that the compact objects are formed through gravitational
collapse and subsequent supernova (SN) explosions, there should be
transient events generating binary compact objects observable by
electromagnetic waves. SN surveys by currently working facilities
e.g., The Subaru Hyper Suprime-Cam (HSC; \citealt{miya12,tomi14}),
Palomar Transient Factory (PTF; \citealt{rau09,law09}), Catalina
Real-Time Sky Survey (CRTS; \citealt{drak09}), Panoramic Survey
Telescope \& Rapid Response System (Pan-STARRS1; \citealt{kais10}),
and SkyMapper \citep{kell07}, and also by coming future projects
(e.g., Large Synoptic Survey
Telescope\footnote{http://www.lsst.org/lsst/}; LSST) will be able to
give constraints on the formation rate of transient objects including
binary compact objects.

One of the possible candidates for a SN forming a close binary system
is {\it ultra-stripped} SN \citep{taur15}, which would launch peculiar
type Ib/c SN with a faint and fast decaying light curve. Peak
luminosity of Type Ib/c SNe is determined primarily by the ejected
mass of $^{56}$Ni, $M_{^{56}\mathrm{Ni}}$, while the timescale around
the peak is determined by the diffusion timescale $\tau_c\propto
M_\mathrm{ej}^{3/4}E_K^{-1/4}$, where $M_\mathrm{ej}$ is the ejecta
mass and $E_K$ is the kinetic energy of the ejecta \citep{arne82}.
Therefore, the low peak luminosity and short characteristic time imply
the small masses of the ejecta and $^{56}$Ni.
For instance, type Ic SN 2005ek is one of these SNe
\citep{drou13,taur13}, whose estimated ejecta mass, $\sim
O(0.1)M_\odot$ is notably smaller than typical SN Ic, $O(1)M_\odot$
\citep{drou11}, as well as smaller $^{56}$Ni mass, and the explosion
energy is also smaller by an order of magnitude ($O(10^{50})$ erg)
than typical core-collapse SNe ($O(10^{51})$ erg).  To model these
rapidly-evolving SNe with small ejecta mass, the progenitor stars are
thought to be stripped much more than canonical stripped-envelope type
Ib/c SNe, that is, ultra-stripped SNe coined by \cite{taur13,taur15}.
Besides SN 2005ek and other known SNe, ten rapidly-evolving transients
were recently detected by Pan-STARRS1, which exhibit shorter decaying
timescale ($\sim 10$ days) than canonical SNe with peak barometric
luminosities ranging from $\sim 10^{42}$ to $10^{43}$ erg s$^{-1}$
\citep{drou14}.
These ultra-stripped SNe are conjectured as products of close binary
systems that experienced strong binary interactions, e.g. common
envelope phase, which result in close binary compact
objects. Therefore, by assuming that the order of the ultra-stripped
SN rate is the same as that of the merger rate, we will be able to
measure the merger rate through SN surveys.
The current predictions of NS merger rate is between $\sim 10^{-6}$
and $4\times 10^{-3}$ galaxy$^{-1}$ year$^{-1}$ \citep{abad10}, which
are corresponding to $\sim 0.01$\% and 40\% of core-collapse SN rate
($\sim$ 0.01 galaxy$^{-1}$ year$^{-1}$), while observed ultra-stripped
SN rate would be $\sim$ 0.1-1\% of supernovae \citep{taur15}.
After the detection of GWs, this conjecture can be tested using
statistics of GW sources.

To predict the properties of these transient objects for coming LSST
era, we investigate the explosion characteristics of the
ultra-stripped SNe by means of numerical simulations. The current
standard model of the explosion mechanism for core-collapse SNe is
based on a neutrino-driven delayed explosion scenario
\citep{colg66,beth85}, in which copious amount of neutrinos emitted in
the vicinity of a newly-born NS are absorbed by the surrounding
materials and they effectively act as heating \citep[see][for latest
  reviews and references therein]{jank12,kota12,burr13,fogl15}. In
this study, we perform i) stellar evolutionary simulations of bare
carbon and oxygen (CO) cores without massive hydrogen and helium
envelopes, which would be possible consequences of close binary
interactions such as common envelope phase, until iron cores form and
ii) axisymmetric neutrino-radiation hydrodynamics simulations to
investigate their explosions.
The paper is organized as follows. Section \ref{sec:progenitor}
describes our stellar evolutionary simulations and stellar structure
just prior to the collapse. The numerical method of following
hydrodynamics simulations and the results are presented in Section
\ref{sec:hydro}. We summarize our results and discuss their
implications in Section \ref{sec:summary}.

\section{Stellar evolution and progenitor structures}
\label{sec:progenitor}

Here, we describe the evolution of CO cores obtained in this study.
To make initial conditions for hydrodynamics simulations, we first
perform stellar evolutionary simulations of CO cores with masses of
1.45, 1.5, 1.6, 1.8, and 2.0 $M_\odot$ supposing that stellar mass
loss has already occurred by their hypothetical companion NSs. By
removing stellar envelope, the stellar evolutionary simulations are
done with a code described in \cite{umed12,taka13,yosh14}. The nuclear
reaction network consists of 300 species of nuclei
\citep{taka13,yosh14}. Schwarzschild criterion is employed as the
convection criterion and the convective mixing of the chemical
composition is evolved using diffusion equation. We also take into
account thermohaline mixing \cite[e.g.][]{sies09} and the diffusion
coefficient is adopted from Eq. (2) in \cite{sies09}. The initial
chemical compositions of the CO cores are evaluated by those after the
He burning with the metallicity $Z=0.02$. Since the C/O ratio of the
core depends on the stellar mass, we assume the mass fractions of
$^{12}$C and $^{16}$O, which are listed in Table \ref{tab}.

It is quite difficult to relate between ZAMS mass and CO-core mass in
binary systems because mass transfer proceeds very complicatedly
during the evolution.  On He star-NS binaries, He stars with
$M_\mathrm{He} \la 3.5 M_\odot$ and with the initial orbital period of
$P_\mathrm{orb} \la 0.5\mathrm{d}$ have an evolutional path to
ultra-stripped SNe \citep{taur15}.  These He stars evolved to CO cores
with $M_\mathrm{CO} \la 1.8 M_\odot$ surrounded by very thin He
envelope.  Thus, the mass range of our CO core models is adequate for
the progenitors of ultra-stripped SNe.  For comparison, we list the
stellar mass at the zero-age main-sequence (ZAMS), $M_\mathrm{ZAMS}$,
which makes the similar CO core as shown in Table \ref{tab}.  The mass
range of the He cores is 2.6 - 3.3 $M_\odot$.  These values are
evaluated based on the evolution of 9 - 15 $M_\odot$ stars up to the
Ne ignition or the central C exhaustion \citep[see also][]{woos15}.
We assume the convective overshooting as a diffusion process until the
core He burning in these calculations. The diffusion coefficient is
adopted from Eq. (2) in \citet{herw00} and the parameter on the scale
height $f$ is set to be 0.01.

\begin{figure}
\centering
\includegraphics[angle=270,width=0.45\textwidth]{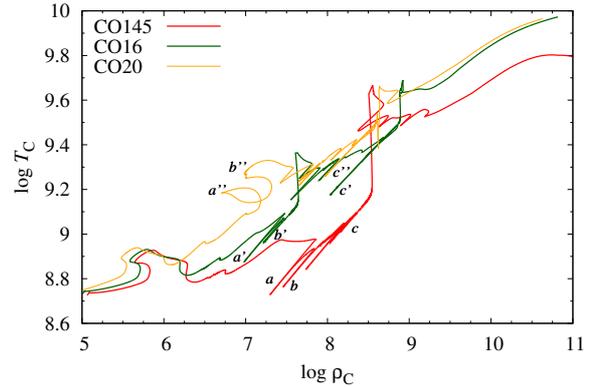}
\caption{Evolutionary path in the plane of the central density and
  temperature of CO145 (red line), CO16 (green line), and CO20 (orange
  line) models.  Labels $a$ and $a'$, $b$ and $b'$, and $c, c',$ and
  $c''$ correspond to the off-center Ne, O, and Si burnings,
  respectively.  Labels $a''$ and $b''$ denote the central Ne and O
  burnings.  }
\label{fig:rhoctc}
\end{figure}

\begin{figure}
\centering
\includegraphics[angle=270,width=0.45\textwidth]{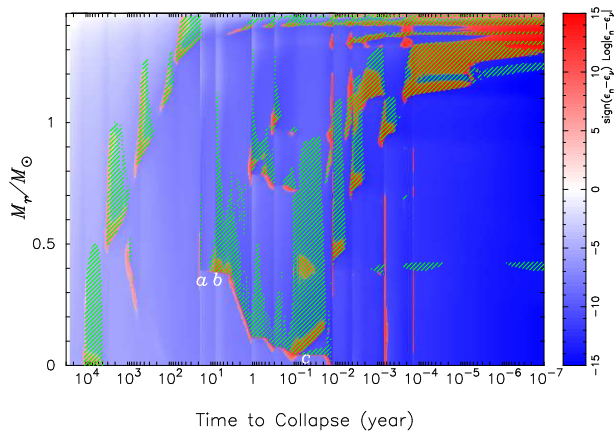}
\caption{Kippenhahn diagram of CO145 model.  Green-hatched areas
  indicate convective regions. Labels $a$, $b$, and $c$ correspond to
  the off-center Ne, O, and Si-burnings at 17.1 years, 10.7 years, and
  38.8 days before core-collapse.  }
\label{fig:HR-CO145}
\end{figure}

\begin{figure*}
\centering
\includegraphics[angle=270,width=0.45\textwidth]{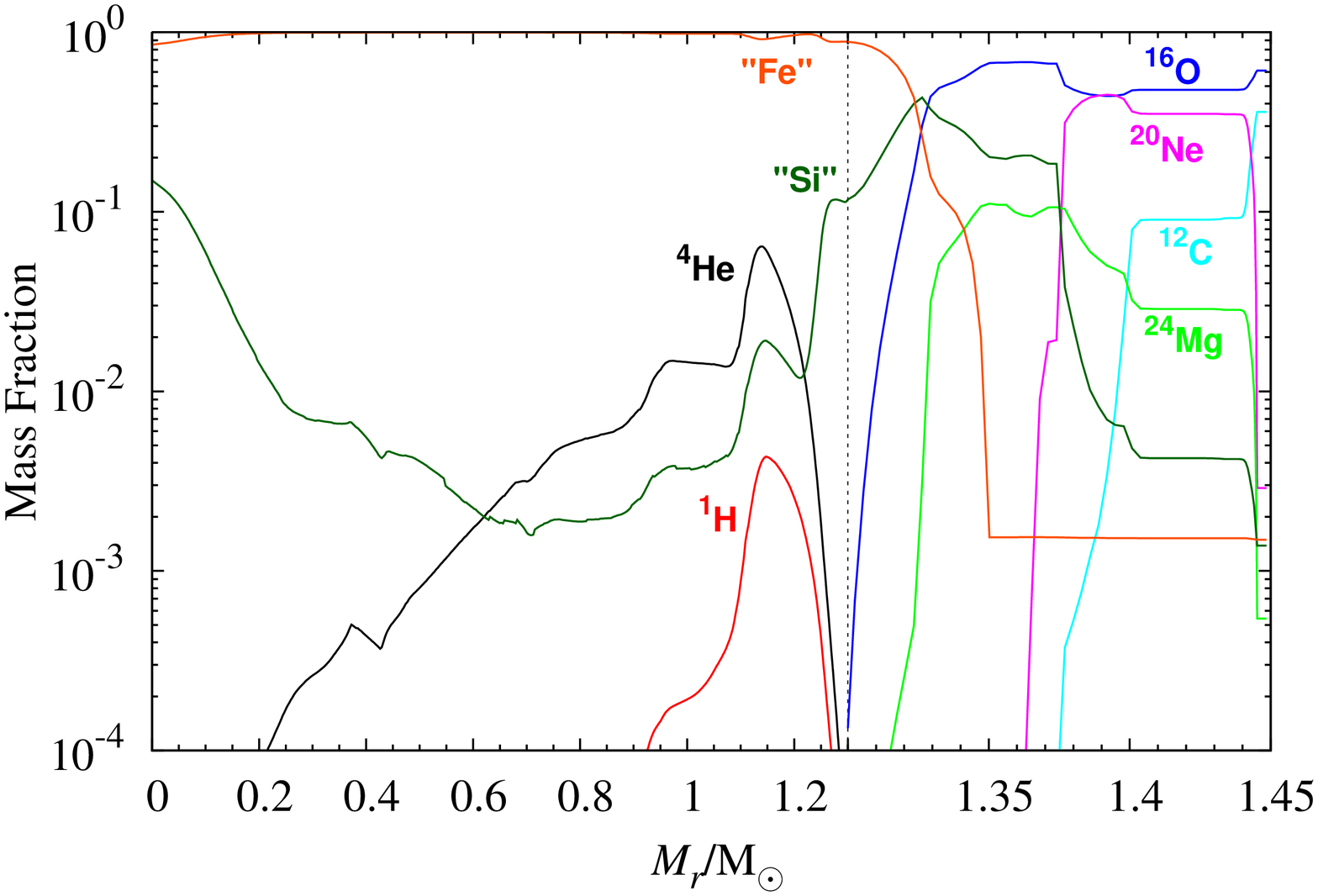}
\includegraphics[angle=270,width=0.45\textwidth]{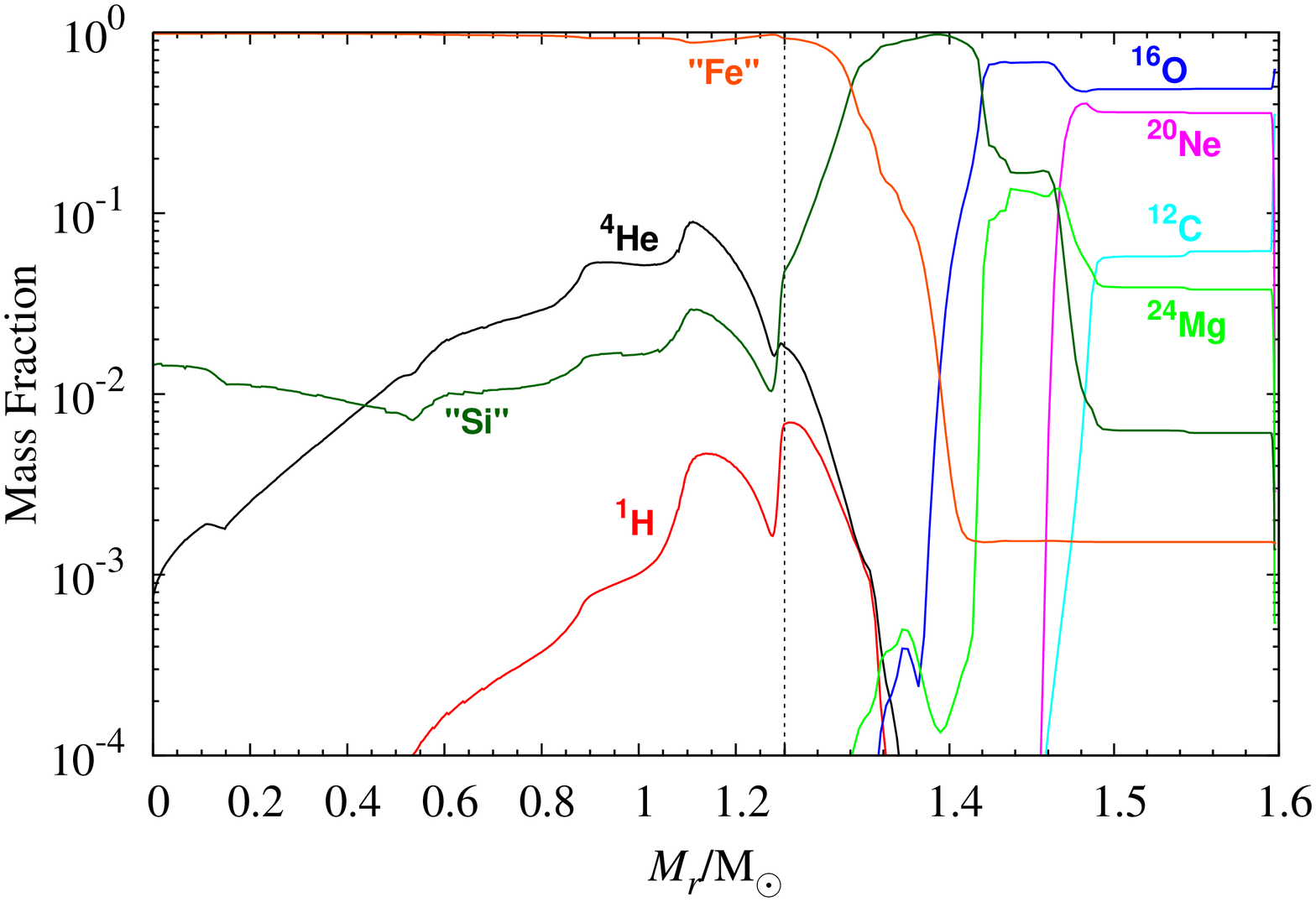}
\caption{Mass fraction distributions of CO145 (left) and CO16 (right)
  models. Red, black, sky-blue, blue, pink, yellow-green, green, and
  orange lines correspond to $^1$H, $^4$He, $^{12}$C, $^{16}$O,
  $^{20}$Ne, $^{24}$Mg, ``Si", and ``Fe".  ``Si" and ``Fe" indicate
  the element groups in $Z = 14 - 21$ (Si-Sc) and $Z \ge 22$ (Ti-),
  respectively.  The scale of the horizontal axis changes at 1.3
  M$_\odot$.}
\label{fig:mfdist}
\end{figure*}

We continue evolutionary simulations until the iron cores become
unstable and start collapsing.  Here we briefly explain the evolution
of our models.  More details will be shown in Yoshida et al. (2015, in
preparation).  Figure \ref{fig:rhoctc} shows the evolution of the
central temperature ($T_C$) and density ($\rho_C$) of CO145, CO16, and
CO20 models. Here, the units of $T_C$ and $\rho_C$ are Kelvin and g
cm$^{-3}$, respectively. Figure \ref{fig:HR-CO145} is the Kippenhahn
diagram of CO145 model, which displays the time evolution of
convective regions and the energy generation inside the star.
Although all models finally form Fe cores, the evolution paths depend
on the CO-core mass.  In CO145 model, off-center Ne burning is ignited
(Ne-flash) after several C-shell burnings (label $a$ in Figures
\ref{fig:rhoctc} and \ref{fig:HR-CO145}).  The following O-burning is
also ignited at an off-center region (label $b$).  Then, the burning
front gradually moves into the center and the temperature at the front
increases.  When the central density reaches $\log \rho_{\rm C} \sim
8.2$, Si-burning is ignited at the burning front and the convective
layer extends to $M_r \sim 0.9 M_\odot$ (label $c$).  The burning
front continues to move inwards and the O/Ne-enriched material in the
central region is burned to Fe-peak elements.  When the central
density reaches $\log \rho_{\rm C} \sim 8.5$, the burning front
reaches the center and the central temperature raises steeply.  The Fe
core grows up to $\sim 0.5 M_\odot$ after the burning front reaches
the center.  Then, shell Si-burning occurs and the Fe core grows up
further, and finally, the core collapses.

CO16 model evolves on a different evolution track as shown in Figure
\ref{fig:rhoctc}.  After the Ne flash (label $a'$) and off-center O
burning (label $b'$), the burning front reaches the center.  As a
result, the Si core with 0.6 $M_\odot$ is formed and it grows up by
O-shell burnings.  Then, the Si flash occurs at $M_r = 0.03 M_\odot$
(label $c'$).  The Si flash forms large convective region up to $M_r
\sim 1.1 M_\odot$ above the burning front.  After that, Si-shell
burnings also follow.  The burning front again gradually moves inwards
and finally reaches the center after the second Si-shell burning.
Finally, the Fe core grows to 1.34 $M_\odot$ and collapses.  The
evolution of CO15 model is similar to that of CO16.  The Ne-flash and
Si-flash occur at higher central density.

The evolution of CO20 model is similar to those of massive stars that
collapse to normal SNe until O-shell burning. Ne and O burnings are
ignited at the center and the convective core forms (labels $a''$ and
$b''$).  The Ne and O-shell burnings extend the central Si core after
O-core burning.  However, Si burning starts at an off-center region
when the central density reaches $\log \rho_{\rm C} \sim 8.6$ (label
$c''$).  The off-center Si burning forms a large convective region to
1.2 $M_\odot$.  The burning front moves inwards and reaches the center
during Si-shell burning.  Finally, an 1.37 $M_\odot$ Fe core is
formed, and then, it collapses. The evolution of CO18 model is similar
to CO20 model, i.e., Ne and O are ignited at the center and Si ignites
at an off-center region.  The central density at the ignition of
off-center Si-burning is higher than that of CO20.

We note that recently the evolution of 9 - 11 $M_\odot$ stars has been
investigated in \cite{woos15}.  Since they considered artificial
energy deposition to the cooler underlying zone of the off-center
burning front, the burning front rapidly moved inward.  We do not
consider this effect, but we instead set larger coefficient value of
thermohaline mixing.  In CO145 model, the central contraction makes
the temperature at the burning front high enough to ignite Si before
the front reaches the center.  Even so, CO145 model forms an Fe core.
Although the inward motion of the burning front depends on the
instability at the base of the burning front and has uncertainty, we
expect that the stars igniting Ne-flash will form an Fe core and
collapse.

The mass fraction distributions of CO145 and CO16 models at the final
step of the stellar evolution simulations are shown in Figure
\ref{fig:mfdist}.  These models consist of a 1.33 and 1.34 $M_\odot$
Fe cores surrounded by thin Si, O/Si, O/Ne, and O/C layers.
Composition-inverted layers have been seen during the Ne-flash and Si
flash.  The composition inversion has been removed during the inward
motion of the burning front.  We do not see large differences in Fe
core masses in our CO core models (see Table \ref{tab}).

We assume the mass fractions of C and O in the CO cores based on the
results of single star models.  In the case of binary evolution,
ultra-stripped CO cores have lost the H-rich envelope before or during
the early phase of the He-core burning.  Stripping the H-rich envelope
prevents the growth of He core and shortens the time scale of the
He-core burning.  Thus, the C/O ratio of an ultra-stripped CO core is
larger than the CO core in single stars \citep{brow01}.
\citet{well99} investigated the evolution of various binary systems of
massive stars.  They obtained that a binary system consisting of 13
$M_\odot$ and 12 $M_\odot$ stars with an initial orbital period of 3.1
days experiences Case B+BB mass transfer and the primary star becomes
a 1.42 $M_\odot$ He star with CO core mass being 1.31 $M_\odot$ and
the central carbon mass fraction being 0.40.

We calculate the evolution of 1.45 and 1.6 $M_\odot$ CO core models
with a large C/O ratio to investigate the C/O-ratio dependence of
ultra-stripped SN progenitors.  We set the mass fractions of C and O
in these models to be 0.471 and 0.500, respectively.  Properties of
the models are shown as CO145c and CO16c in Table \ref{tab}.  We do
not see large systematic differences between normal and large
C/O-ratio models.  CO145c model has an Fe core slightly smaller than
CO145, while the Fe core of CO16c is slightly larger than that of CO16
for the criterion that the Fe core is determined by the electron
fraction smaller than 0.495.  The difference of the compactness
parameter between the normal model and large C/O-ratio model is within
10 \%.  We consider that properties of ultra-stripped SN progenitor do
not strongly depend on the C/O ratio.
It should be noted that the evolution of the central region of the CO
cores depends on the C/O ratio.  The convective regions of the central
or shell C-burnings become large in the models of the large C/O ratio.
In CO145c model, Ne is ignited at 0.35 $M_\odot$ in the mass
coordinates.  The Ne/O-burning front reaches the center before the Si
ignition.  Then, Si is ignited at 0.04 $M_\odot$.  In CO16c model,
Si-burning occurs at the center as a flash and, then, it turns to
steady burning.  Nevertheless, we do not see large difference by the
C/O ratio in the region where the structures and composition are
mainly determined by the Si-shell burning.  Once the Fe core is formed
by the central or off-center Si-burning, the core grows through the
following Si-shell burning.  Thus, we consider that properties of SN
progenitors such as Fe core mass and compactness parameter are
insensitive to the C/O ratio.

These CO cores have a thin CO envelope where shell C burnings scarcely
affect the composition.  We listed the masses of the envelope,
$M_\mathrm{env}$, which the convective region did not reach during
C-shell burnings, and their binding energy, $E_\mathrm{bind,env}$, in
Table \ref{tab}. The envelope mass is less than 0.01 $M_\odot$ for all
the models.  The binding energy of the envelope is $(0.4 - 2) \times
10^{48}$ erg, which is smaller than one per cent of the binding energy
of the whole star ($E_\mathrm{bind}$, see Table \ref{tab}).

Although we do not take into account He-rich envelope, the progenitors
of ultra-stripped SNe in NS binary systems could have the envelope
\citep{taur13, taur15}.
\cite{taur15} showed the dependence of the He mass and the binding
energy of the He-envelope on the final CO core mass.  Here, we
estimate the binding energy of the He-envelope from Figure 15 in
\cite{taur15} and from their discussion that the He mass is smaller
than 0.2 M$_\odot$ \citep[see \S 4.1.1 in][]{taur15}.  The estimated
range of the envelope is listed in Table \ref{tab}.  The binding
energy of the He envelope is at most several times $10^{49}$ erg.
Thus, we expect that the envelope has an small contribution in the
binding energy for ultra-stripped SNe.

\begin{figure}
\centering
\includegraphics[width=0.45\textwidth]{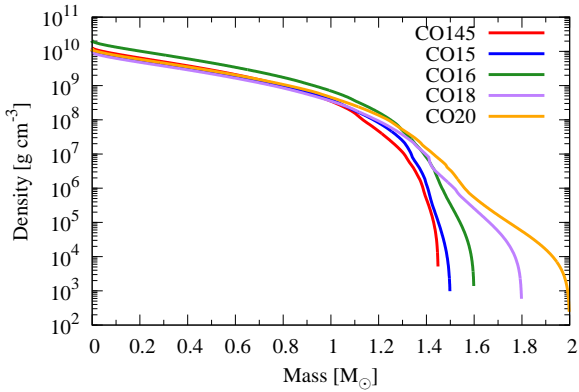}
\caption{Density structure of CO cores at precollapse phase as a
  function of mass coordinate. Red, blue, green, purple, and orange
  lines represent models with CO cores of 1.45, 1.5, 1.6, 1.8, and
  2.0$M_\odot$, respectively.  }
\label{fig:density}
\end{figure}

Figure \ref{fig:density} presents density structures at the end of
evolutionary simulations. More detailed properties are summarized in
Table \ref{tab}. Following \cite{ocon11}, we define the compactness
parameter, $\xi_M$, by
\begin{equation}
\xi_M=\frac{M/M_\odot}{R(M)/1000~\mathrm{km}},
\end{equation}
where $R(M)$ is the radius with $M$ being mass enclosed inside $R$. We
employ $\xi_{1.4}$, which is measured by $M_r=1.4M_\odot$.  Note that
other groups have employed different values such as $\xi_{2.5}$,
$\xi_{1.75}$, $\xi_{1.5}$
\citep{ocon11,ugli12,ocon13,sukh14,naka14,pejc15}. Since our models
have small CO core mass, we rely on a different parameter.
Note also that in the original definition by \cite{ocon11}, $\xi$ is
measured at the time of bounce. Later, \cite{sukh14} suggested that
$\xi$ measured just prior to collapse correlates with $\xi$ at
bounce. Hence, in this paper we evaluate $\xi$ with the precollapse
density structures.
We show two different definitions of the iron core mass,
$M_\mathrm{F_e}$: one is based on the electron fraction and the other
on mass fraction of heavy elements.  The iron core mass does not
depend strongly on CO core mass since, broadly speaking, the core
collapse sets in when the iron core mass exceeds Chandrasekhar mass
\citep{baro90}.

\begin{table*}
\centering
\caption{Properties of progenitor models}
\begin{tabular}{lcccccccccccccccc}
\hline
Model &
$M_\mathrm{CO}$ &
$M_\mathrm{ZAMS}$ $^a$ &
$X_{\rm C}$ &
$X_{\rm O}$ &
Radius &
$\xi_{1.4}$ &
$M_\mathrm{F_e}$ $^b$ &
$M_\mathrm{F_e}$ $^c$ &
$M_\mathrm{env}$ &
$E_\mathrm{bind,env}$ &
$E_\mathrm{bind,He}^d$ &
$E_\mathrm{bind}$
\\
 & [$M_\odot$] & 
[$M_\odot$] & & & 
[$10^4$km] & &
[$M_\odot$] & 
[$M_\odot$] & 
[$10^{-3} M_\odot$] &
[$10^{48}$ erg] &
$10^{49}$ erg] &
[$10^{50}$ erg]
\\
\hline
CO145 & 1.45 & 9.75 & 0.360 & 0.611 & 1.28 & 0.468 & 1.33 & 1.32 & 4.80 & 2.28 & 0.1--1.3 & 3.97 \\
CO145c & 1.45 & 9.75 & 0.471 & 0.500 & 1.78 & 0.502 & 1.31 & 1.28 & 1.45 & 0.46 & 0.1--1.3 & 3.99 \\
CO15 & 1.5 & 10.0 & 0.360 & 0.611 & 2.20 & 0.600 &1.31 & 1.29 & 1.60 & 0.38 & 0.5--1.7 & 4.17 \\
CO16 & 1.6 & 10.5 & 0.350 & 0.621 & 1.93 & 0.912 & 1.34 & 1.34 & 2.63 & 0.80 & 0.7--2.8 & 5.39 \\
CO16c & 1.6 & 10.5 & 0.471 & 0.500 & 1.73 & 0.850 & 1.36 & 1.34 & 2.42 & 0.95 & 0.7--2.8 & 4.69 \\
CO18 & 1.8 & 11.5 & 0.350 & 0.621 & 2.64 & 0.851 & 1.35 & 1.37 & 8.09 & 2.20 & 1.2--3.8 & 4.92 \\
CO20 & 2.0 & 12.8 & 0.330 & 0.641 & 3.66 & 0.968 & 1.37 & 1.37 & 8.48 & 1.66 & 2.2--4.7 & 5.72 \\
\hline
\end{tabular}
\begin{flushleft}
$^a$ This is corresponding mass of ZAMS stars, which make the same
  mass of CO core. \\
$^b$ The iron core mass is determined by the mass with the electron
  fraction of $Y_e<0.495$. \\
$^c$ The iron core mass is determined by the mass with the mass
  fraction of the element groups in $Z \ge 22$ (Ti-) larger than 0.5.\\
$^d$ These values are taken from Figure 15 in \citet{taur15}.
\end{flushleft}
\label{tab}
\end{table*}

\section{Hydrodynamics simulations}
\label{sec:hydro}

\subsection{Numerical methods}

For the hydrodynamics simulations, we employ a two-dimensional (2D)
neutrino-radiation hydrodynamics code, which is developed and used for
investigating SN explosion mechanism
\citep{suwa10,suwa11b,suwa13b,suwa14c,suwa14a}. With ZEUS-2D code
\citep{ston92} as a base for the solver of hydrodynamics, we employ an
equation of state of \cite{latt91} with an incompressibility $K=220$
MeV and solve the spectral transfer of neutrinos by the isotropic
diffusion source approximation (IDSA) scheme \citep{lieb09} that
splits the neutrino distribution function into two components, both of
which are solved by using separate numerical techniques.  Weak
interaction rates are implemented following \cite{brue85}. We solve
transfer of electron-type neutrinos ($\nu_e$) and antineutrinos
($\bar\nu_e$), but heavier leptonic neutrinos are not taken into
account.  In our 2D simulations, axial symmetry is assumed and
``ray-by-ray-plus'' approach is implemented for multi-dimensional
treatment of neutrino transfer \citep{bura06a}. Spherical coordinates
$(r, \theta)$ with logarithmic zoning in the radial direction and
constant zoning in $\theta$ are used.  The simulations are performed
on a grid of 300 radial zones extending up to 5000 km with the
smallest grid width being 1 km at the center and 128 equidistant
angular zones covering $0 < \theta < \pi$ for 2D simulations. For
neutrino transport, we use 20 logarithmically spaced energy bins
ranging from 3 to 300 MeV.

Note that at the current moment there is no complete model for the
explosion mechanism of core-collapse SNe despite the long-lasting
efforts. Recent multi-dimensional simulations of neutrino radiation
hydrodynamics have shown diversity of numerical results
\citep[e.g.][used the same progenitor models and obtained different
  results]{brue13,suwa14c,dole15,mels15b}. In addition, most exploding
simulations exhibited order of magnitude smaller explosion energy than
observation and remnant compact objects often gained mass above the
maximum mass of a NS, which would collapse to BHs instead of NSs.

We speculate, however, that all these implementations could
universally reproduce ultra-stripped SNe. The reason for this
speculation is that typical explosion energy of the ultra-stripped SNe
is smaller ($O(10^{50})$ erg) than canonical SNe ($O(10^{51})$ erg),
that is, stellar structures of progenitors for these different kinds
of SNe are different. The explosion energy is determined naively by
the binding energy of progenitor layers in the vicinity of mass cut
(remnant compact object) so that stars with small binding energy are
possible candidates of the weak explosions.  This hypothesis was
already applied for explanation of type IIn-P SNe like SN 2009kn,
which exhibit narrow emission lines, a short plateau phase in light
curve, and small amount of $^{56}$Ni.
These observational features were reproduced by an electron-capture SN
model, in which the progenitor star consists of an O-Ne-Mg degenerated
core and a very thin envelope, giving the small explosion energy of
$\sim 10^{50}$ erg (\citealt{mori14} based on the explosion simulation
by \citealt{kita06}).
As we will show in the following subsection, we obtain explosion
energy of $O(10^{50})$ erg, which is the same order of magnitude as
the binding energy exterior to a remnant compact object at precollapse
phase. We infer that even if we could perform {\it a realistic
  simulation} that would reproduce canonical explosion energy for
canonical SN progenitors, the results obtained with weakly bound
progenitors will be a weak explosion with its energy $\sim 10^{50}$
erg. Since the explosion energy of such simulations, however, could be
larger than that of ours, it is safe to consider that our results give
at least a lower limit of the explosion energy.

At first, we perform spherically symmetric (1D) simulations up to 5 ms
after the bounce, which is determined by the largest central density,
and 2D simulations follow them. Note that all 1D simulations fail to
explode even when we continue simulations until several hundred
milliseconds after the bounce.\footnote{Note that recent studies of
  full three-dimensional (3D) hydrodynamics simulations showed that 2D
  simulations are more favorable for explosion than those of 3D
  \citep[see e.g.,][]{hank13,taki14} so that we should note that the
  2D simulations would optimize the conditions for successful
  explosions \citep[but see also][]{mels15a}.}

\subsection{Results}

\begin{figure}
\centering
\includegraphics[width=0.45\textwidth]{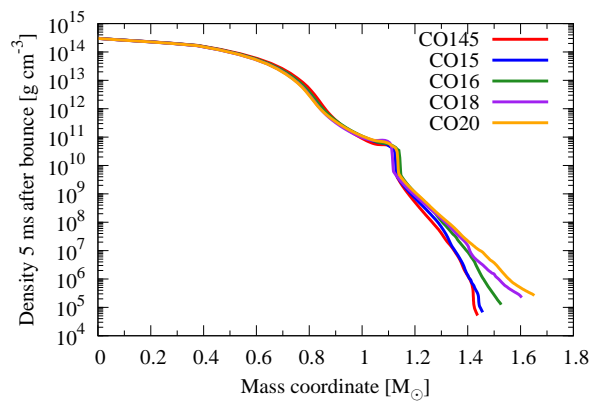}
\caption{Density profiles at 5 ms after bounce. Line colors show the
  same models as Figure \ref{fig:density}.}
\label{fig:d_postbounce5ms}
\end{figure}

We show the density profiles at 5 ms after the bounce in Figure
\ref{fig:d_postbounce5ms}. One can find that the density profiles of
$M_r\lesssim 1.1 M_\odot$ depend only weakly on the CO core mass, but
above this mass coordinate they differ from each other due to the
different progenitor structure. This difference leads to different
evolution of shocks, which is described in the following. Note that
the density at a certain mass coordinate above $\sim 1.1M_\odot$
monotonically increases with the CO core mass.

\begin{figure*}
\centering
\includegraphics[width=0.35\textwidth]{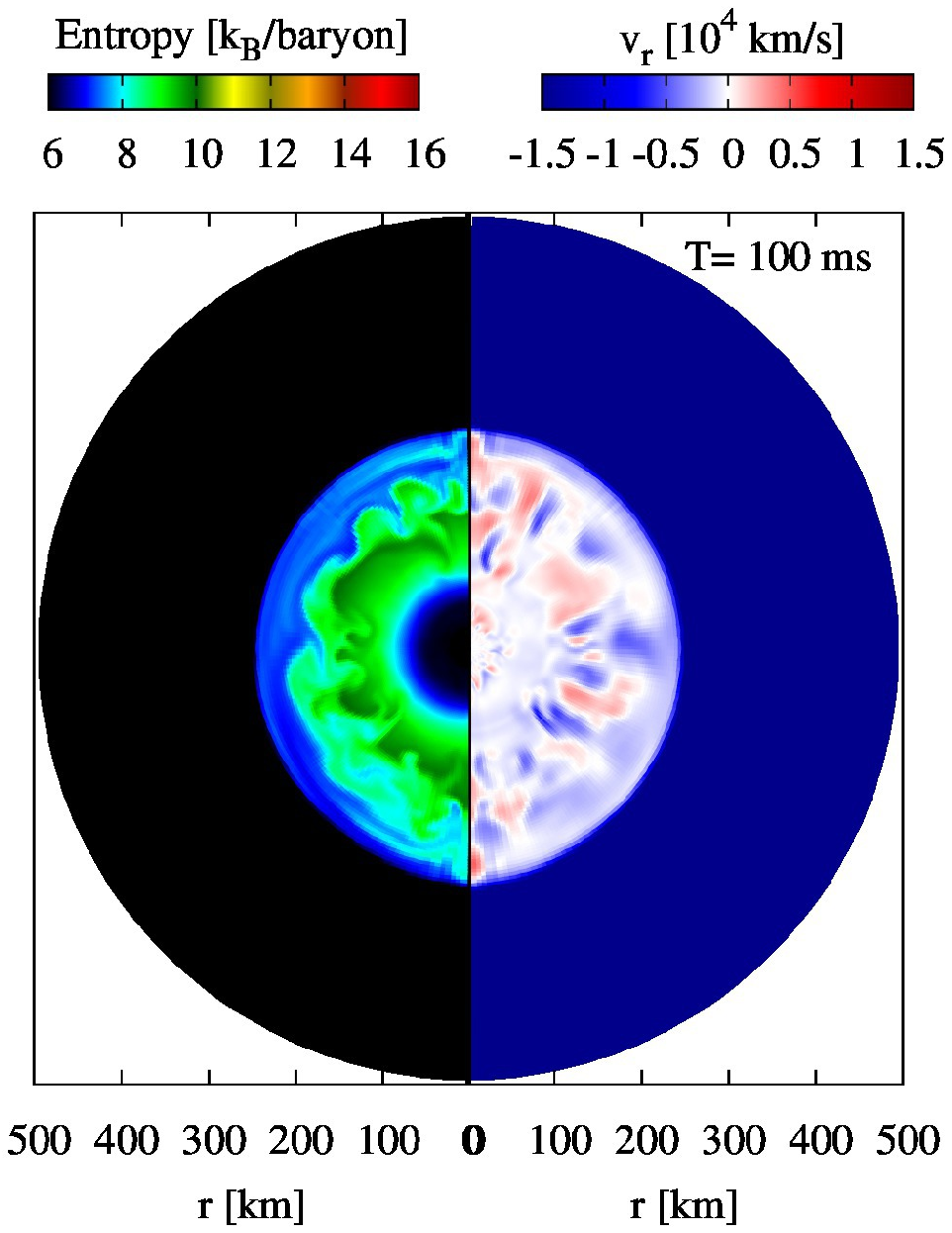}
\includegraphics[width=0.35\textwidth]{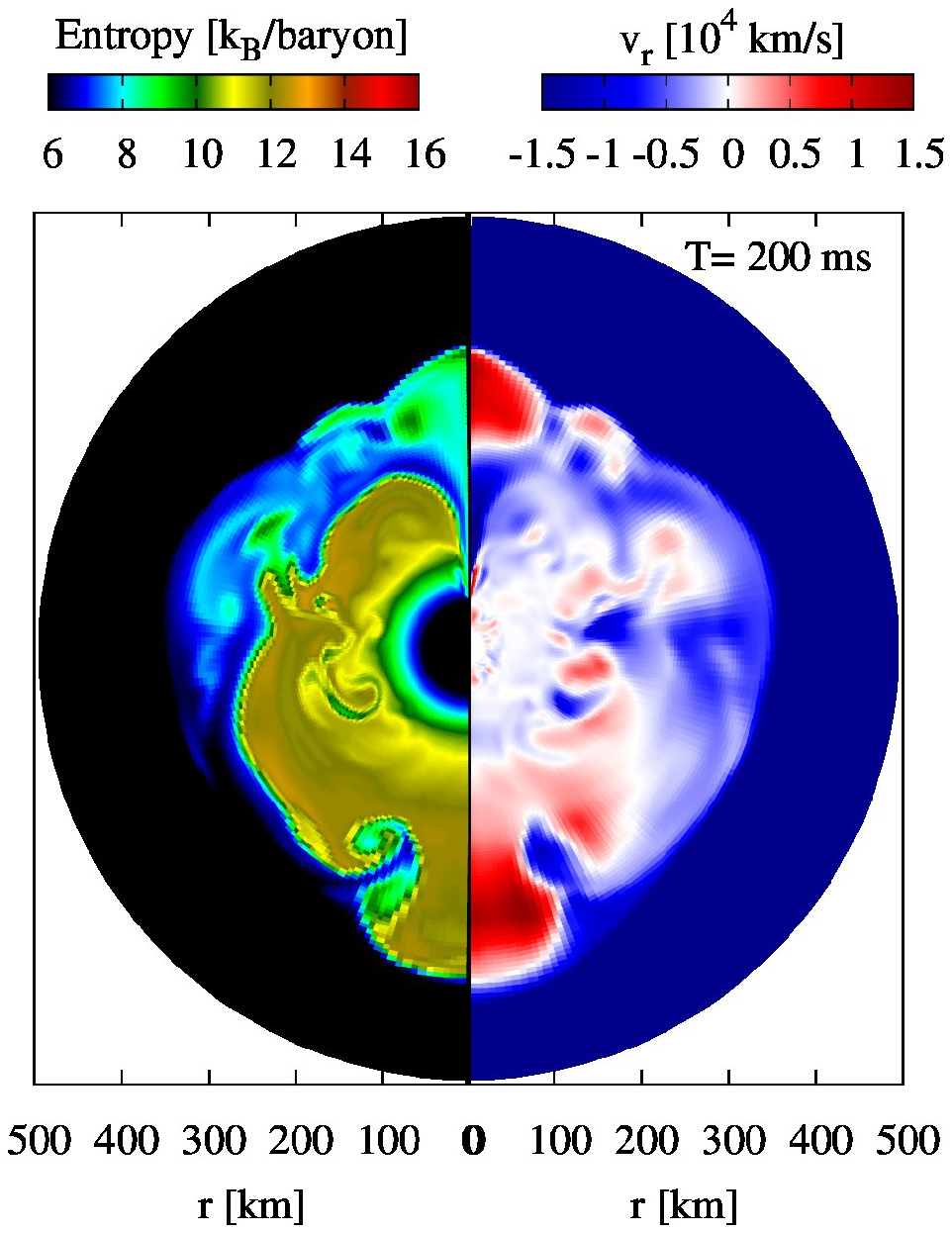}
\includegraphics[width=0.35\textwidth]{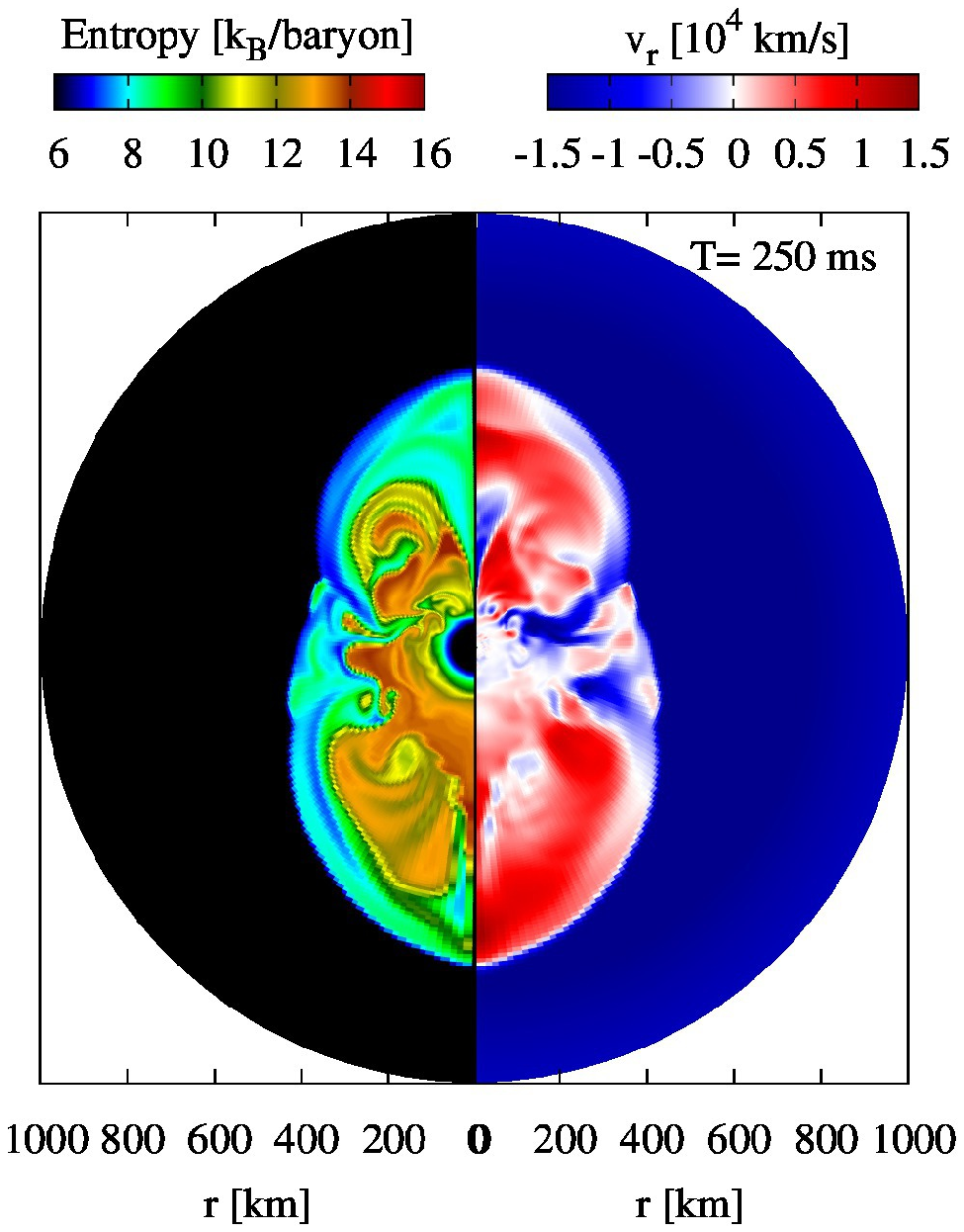}
\includegraphics[width=0.35\textwidth]{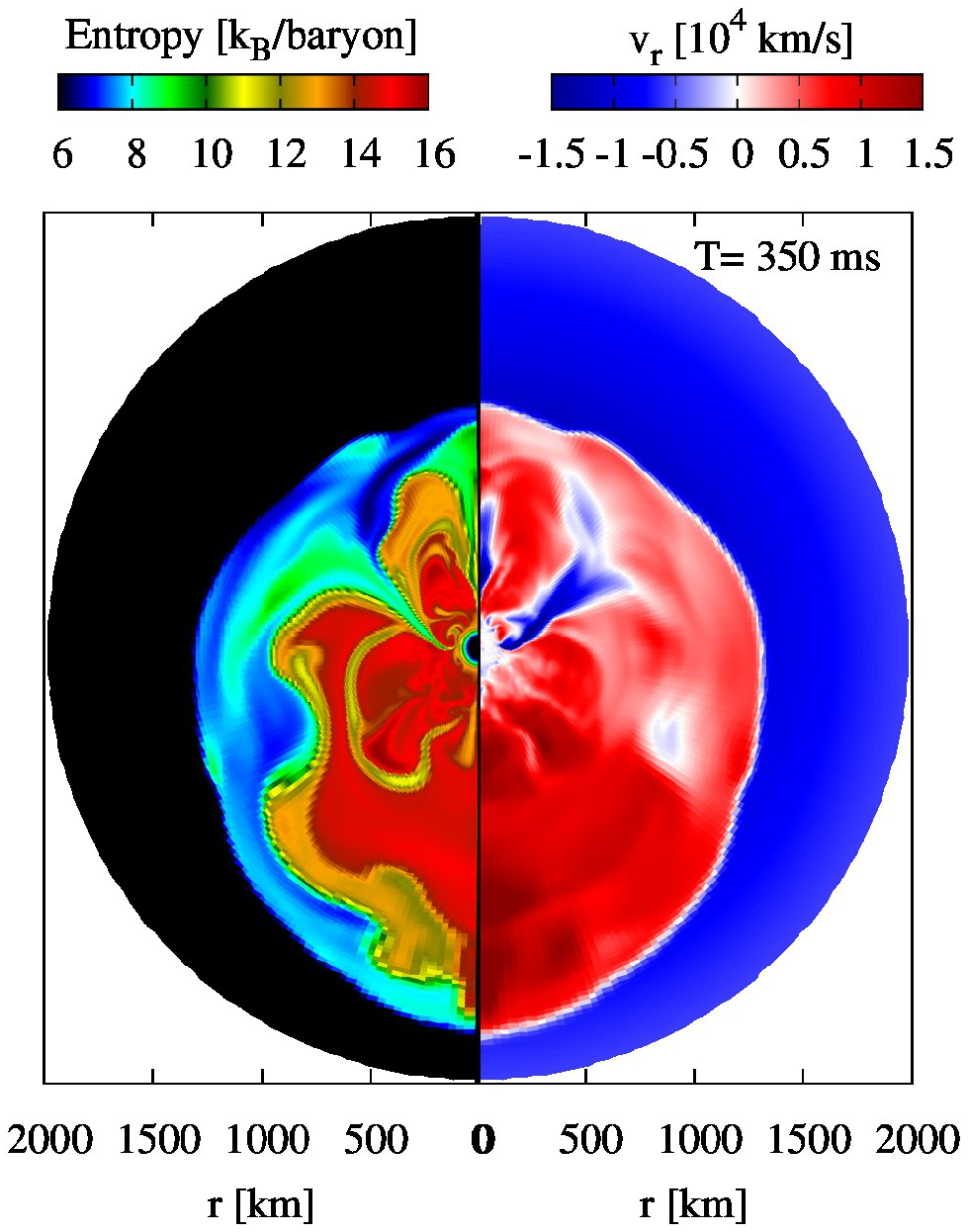}
\caption{Specific entropy (in unit of $k_\mathrm{B}$ baryon$^{-1}$;
  left halves of the individual panels) and radial velocity (in unit
  of $10^4$ km s$^{-1}$; right halves) profiles at 100 (left top
  panel), 200 (right top), 250 (left bottom), and 350 (right bottom)
  ms after the bounce for model CO15. In the entropy plots, blueish
  (reddish) colors represent small (large) entropy. In velocity plots,
  red region is expanding (positive radial velocity) and blue region
  is accreting (negative radial velocity).}
\label{fig:entropy}
\end{figure*}

Figure \ref{fig:entropy} presents evolution of entropy (left halves)
and radial velocity (right halves) distributions at 100 (left top
panel), 200 (right top), 250 (left bottom), and 350 (right bottom) ms
after the bounce for model CO15. After core bounce, convective motion
sets in initially due to an unstable entropy configuration (prompt
convection) and subsequently due to neutrino heating in the gain
region (neutrino-driven convection), but shock morphology is
maintained to be approximately spherically symmetric (see left top
panel). As the turbulent motion is further enhanced by the neutrino
heating, the shock is deformed from spherical symmetry (see right
top). Once the postshock pressure (thermal pressure and turbulent
pressure) overwhelms the ram pressure above the shock, the shock
starts expanding and an explosion is launched. In this model, the
expanding shock is rather spherical since the small envelope mass
results in a small mass accretion rate onto the shock and small-scale
convection dominates over large-scale motion driven by the standing
accretion shock instability \citep[SASI;][]{blon03}. This leads to a
small kick velocity of protoneutron stars (PNS). One can observe a
cold downflow onto the PNS even after the onset of the explosion (see
an isolated blue region in the bottom right panel of Figure
\ref{fig:entropy}), which could increase the PNS mass. This downflow,
however, has a small solid angle and the mass accretion rate is
considerably small. Thus, the mass accreting onto the PNS at this
moment is negligible. The PNS mass evolution will be discussed later.

\begin{figure}
\centering
\includegraphics[width=0.45\textwidth]{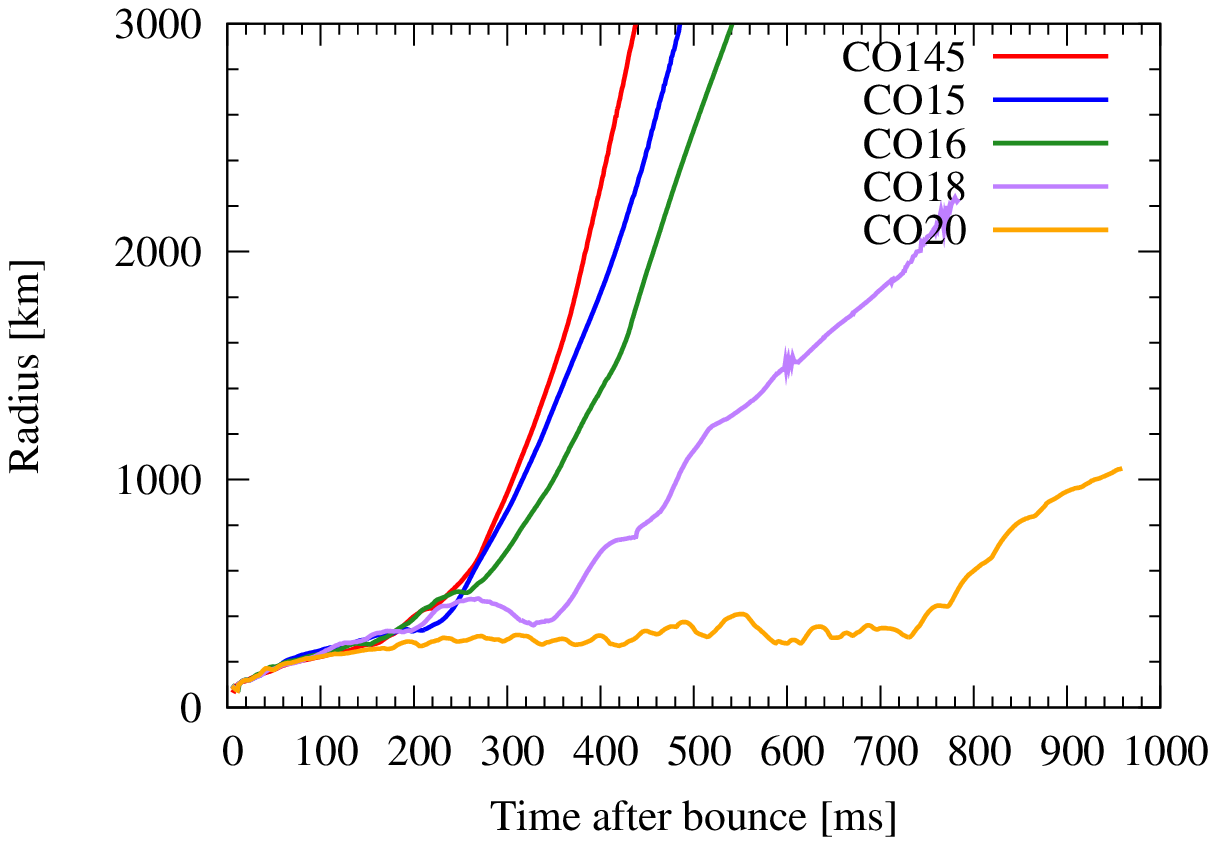}
\caption{Evolutions of the radius of shocks.}
\label{fig:shock}
\end{figure}

\begin{figure}
\centering
\includegraphics[width=0.45\textwidth]{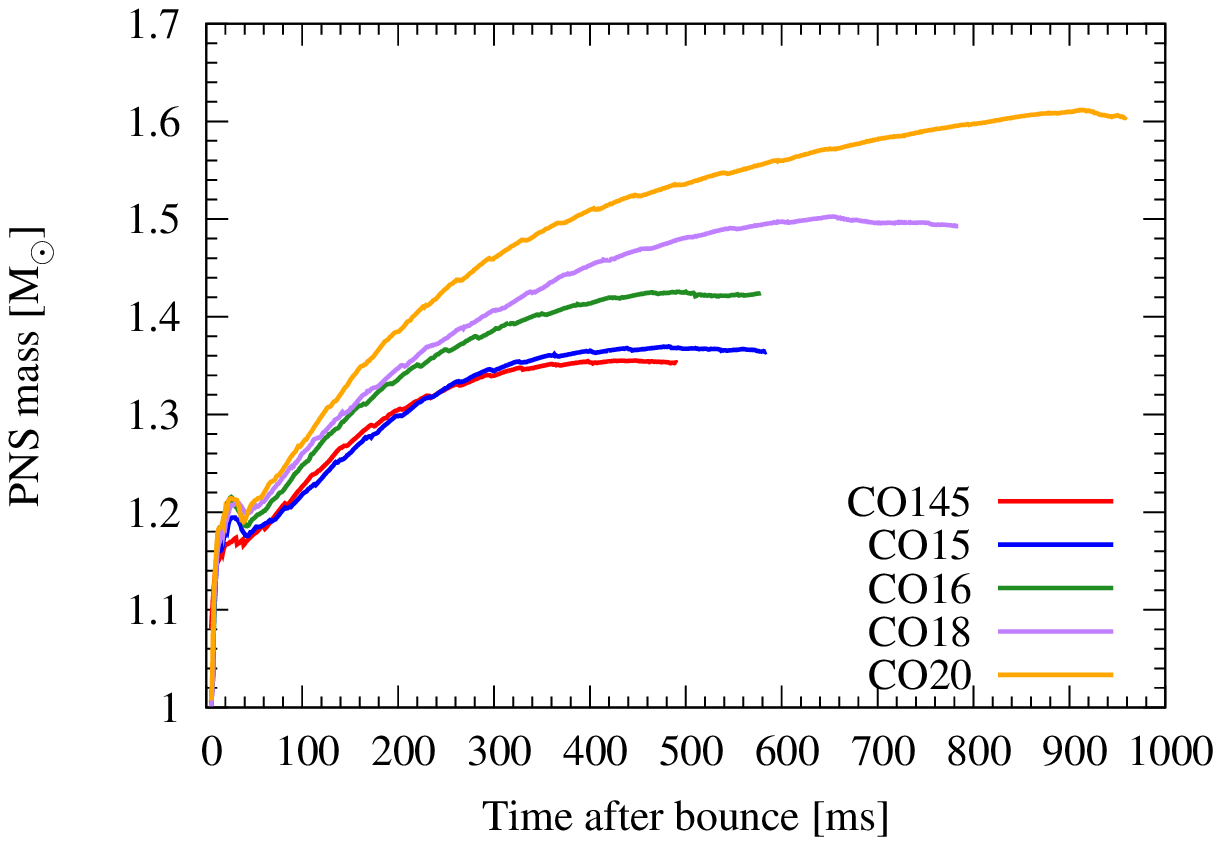}
\caption{Time evolutions of PNS mass (define by $\rho>10^{11}$ g
  cm$^{-3}$).}
\label{fig:mass}
\end{figure}

Shock evolution processes in all 2D simulations are presented in
Figure \ref{fig:shock}. One can see that the three small mass models
(CO145, CO15, and CO16) explode approximately at identical time, while
more massive cores show later explosion. This is a consequence of the
different envelope mass on the iron core. The early exploding models
have clear phase transition from a slowly expanding phase to a rapidly
exploding phase at $\sim$ 200 ms after the bounce. On the other hand,
late exploding models show oscillations of the shock radius before the
explosion. This is due to convection and SASI.  The difference in the
onset time of the explosion is a result of different mass accretion
rate evolution due to the different envelope structure (see Figure
\ref{fig:d_postbounce5ms}). The later onset of the explosion leads to
larger PNS mass as shown in Figure \ref{fig:mass}. Here we define PNS
as the region with density above $10^{11}$ g cm$^{-3}$. CO145 and CO15
models form a PNS of baryonic mass $\approx 1.35 M_\odot$, while other
models give larger PNS mass.

In Table \ref{tab2}, we summarize results of our hydrodynamics
simulations. $t_\mathrm{final}$ denotes the final postbounce time of
each simulation. The quantities listed in other columns are all
measured at $t_\mathrm{final}$. $R_\mathrm{sh}$ is the angle-averaged
shock radius, $E_\mathrm{exp}$ is diagnostic explosion energy, which
is defined as the integral of the sum of specific internal, kinetic
and gravitational energies over all zones with positive value,
$M_\mathrm{NS, baryon}$ is baryonic mass of the remnant NS calculated
by integration over grid of $\rho>10^{11}$ g cm$^{-3}$, $M_\mathrm{NS,
  grav}$ is the corresponding gravitational mass,
$M_\mathrm{ej}=M_\mathrm{CO}-M_\mathrm{NS, baryon}$ is the ejecta
mass, $M_\mathrm{^{56}Ni}$ is mass of $^{56}$Ni, and $v_\mathrm{kick}$
is the estimated kick velocity of NSs. Note that these quantities are
not the final outcome of the simulations, since all the simulations
were terminated before the system relaxes to a stationary state to
save the computational time.
The gravitational mass is calculated by the baryonic mass using the
following relation \citep{timm96}
\begin{equation}
\frac{M_\mathrm{baryon}}{M_\odot}-\frac{M_\mathrm{grav}}{M_\odot}=0.075\left(\frac{M_\mathrm{grav}}{M_\odot}\right)^2.
\label{eq:barionic2gravitational}
\end{equation}
$^{56}$Ni mass is calculated using tracer particle method
\citep[e.g.,][]{naga97}. We assume that the mass elements with the
maximum temperature being over $5\times 10^9$ K achieve nuclear
statistical equilibrium and synthesize $^{56}$Ni completely. This
gives just an approximate estimate. For more realistic calculation we
need to perform detailed nucleosynthesis calculation, which is beyond
the scope of this paper.
The NS kick velocity is estimated by assuming the linear momentum
conservation of the whole progenitor star, i.e., assuming that
anisotropic mass ejection leads to NS kick
\citep[e.g.,][]{wong13}. The linear moment of ejecta is calculated by
\begin{equation}
\mathbf{P}_\mathrm{ej}=\int_{\rho<10^{11}~\mathrm{g~cm^{-3}}, v_r>0}\rho \mathbf{v} \mathrm{d}V,
\end{equation}
where $\mathbf{v}$ is the velocity vector and $v_r$ is its radial
component. The NS kick velocity is then given by
$\mathbf{v}_\mathrm{kick}=-\mathbf{P}_\mathrm{ej}/M_\mathrm{NS,baryon}$.
Since the axial symmetry is assumed in our simulations, the kick
velocity may be overestimated due to the existence of preferable
direction of NS kick, i.e. symmetry axis. Additionally, the stochastic
nature of postshock turbulent flow would also change the degree of
asymmetry of ejecta so that the the initial small perturbation could
change the kick velocity significantly \citep{sche06}. More
statistical study is needed to pin down this issue. It can be argued
that small envelope, not small iron core itself, which can rapidly
accelerate shock, would generally lead to small kick velocity due to
too short time for SASI to build up \citep[see also,
  e.g.,][]{pods04,bogo07}.

\begin{table*}
\centering
\caption{Summary of simulation results}
\begin{tabular}{lccccccccc}
\hline
Model &
$t_\mathrm{final}$$^a$ &
$R_\mathrm{sh}$$^b$ &
$E_\mathrm{exp}$$^c$&
$M_\mathrm{NS, baryon}$$^d$&
$M_\mathrm{NS, grav}$$^e$&
$M_\mathrm{ej}$$^f$&
$M_\mathrm{Ni}$$^g$&
$v_\mathrm{kick}$$^h$&
\\
& 
[ms] &
[km] &
[B] &
[$M_\odot$] &
[$M_\odot$] &
[$10^{-1}M_\odot$] &
[$10^{-2}M_\odot$] &
[km s$^{-1}$] &
\\
\hline
CO145 & 491& 4220 & 0.177 & 1.35 & 1.24 & 0.973 & 3.54 & 3.20\\
CO15 & 584 & 4640 & 0.153 & 1.36 & 1.24 & 1.36 & 3.39 & 75.1\\
CO16 & 578 & 3430 & 0.124 & 1.42 & 1.29 & 1.76 & 2.90 & 47.6\\
CO18 & 784 & 2230 & 0.120 & 1.49 & 1.35 & 3.07 & 2.56 & 36.7\\
CO20$^i$ & 959 & 1050 & 0.0524 & 1.60 & 1.44 & 3.95 & 0.782 & 10.5\\
\hline
\end{tabular}
\begin{flushleft}
$^a$ The final time of simulations measured by postbounce time.\\
$^b$ The angle-averaged shock radius at $t_\mathrm{final}$.\\
$^c$ The explosion energy in unit of B (=$10^{51}$ erg) at $t_\mathrm{final}$, which is still increasing.\\
$^d$ The baryonic mass of NS at $t_\mathrm{final}$.\\
$^e$ The gravitational mass of NS computed by Eq. (\ref{eq:barionic2gravitational}) at $t_\mathrm{final}$.\\
$^f$ The ejecta mass at $t_\mathrm{final}$.\\
$^g$ The Ni mass at $t_\mathrm{final}$.\\
$^h$ The kick velocity at $t_\mathrm{final}$.\\
$^i$ Note that this model is marginally exploding.
\end{flushleft}
\label{tab2}
\end{table*}

\section{Summary and discussion}
\label{sec:summary}

We have performed both stellar evolution simulations of bare CO cores
and explosion simulations for the end product of the CO cores for
modeling ultra-stripped type Ic SNe. We have found that all CO cores
with mass from 1.45 to 2.0 $M_\odot$ resulted in explosion with energy
of $O(10^{50})$ erg, which left NSs with gravitational mass from
$\sim$ 1.24 to 1.44 $M_\odot$ and ejecta from $\sim 0.1$ to 0.4
$M_\odot$ with synthesized $^{56}$Ni of $O(10^{-2}) M_\odot$. These
values are compatible with observations of ultra-stripped SN
candidates \citep{drou13,taur13,taur15}. For SN 2005ek,
$M_\mathrm{ej}\approx$ 0.2--0.7 $M_\odot$ and $M_\mathrm{Ni}\approx$
0.02--0.05 $M_\odot$ are appropriate to fit its light curve. The event
rate of these SNe is estimated as $\sim 1$\% of core-collapse SN rate
\citep{drou13,drou14}, which is also compatible with an NS merger rate
estimation \citep{abad10}.

We took a different approach from previous studies on ultra-stripped
SNe \citep{taur13,taur15}. In previous works, they self-consistently
performed stellar evolutionary simulations until oxygen burning phase
with self-consistent mass loss driven by wind but explosion
calculations were based on phenomenological modeling with three free
parameters; kinetic energy of SN, Ni mass, and mass cut (i.e. NS
mass). Based on this model, they found that ultra-stripped SN model
could account for the light curve of SN 2005ek quite well.
In our work, on the other hand, we performed stellar evolutionary
simulations until the last phase of evolution, i.e., iron core
collapse, but for initially bare CO cores without mass loss. For the
explosion phase, we performed neutrino radiation hydrodynamics
simulations to calculate explosion energy, Ni mass, and NS baryon mass
in self-consistent manner. In this sense, this work is complementary
to previous works. We found that outcomes of our hydrodynamics
simulations for CO145, CO15, and CO16 are close to model parameters of
\cite{taur13} that well fit light curve of SN 2005ek
($M_\mathrm{NS}=1.3M_\odot$, $M_\mathrm{ej}=0.2M_\odot$,
$E_\mathrm{exp}=5\times 10^{50}$ erg, and
$M_\mathrm{Ni}=0.05M_\odot$). The light curve calculated with our
hydrodynamical models will be presented in the forthcoming paper.

For CO145, CO15, CO16, and CO18 models, the baryon mass of the remnant
NSs is in the range between $1.35M_\odot$ and $1.49M_\odot$.  The
corresponding gravitational mass are 1.24 - $1.35M_\odot$ for this
baryon mass range depending only weakly on equations of state for
neutron stars hypothetically employed.  These values for the
gravitational mass agree well with those for secondary NSs in observed
compact binary systems \citep{latt12b}. This suggests that such
secondary NSs may be formed from ultra-stripped SNe.

We showed that $1.45 - 2.0 M_\odot$ CO star models ignite Ne at the
center or an off-center region. They form an Fe core and none of them
evolve to electron-capture SN. Thus, the upper-limit of the CO core
mass for electron-capture SN should be less than 1.45 M$_\odot$.
\citet{taur15} adopted $M_\mathrm{ONe, f} = 1.43 M_\odot$ as an
approximate upper limit for electron-capture SN.  In the case of
single star evolution in \citet{taka13}, a 10.8 $M_\odot$ model ends
its evolution as an electron-capture SN and an 11.0 $M_\odot$ model
ignites Ne at an off-center region.  Off-center Ne ignition and
gradual increase in the central temperature around the central density
$\rho_\mathrm{C} \sim 10^9$ g cm$^{-3}$ could be predictions of the Fe
core formation and core-collapse SN.  The evolution of CO cores less
massive than 1.45 $M_\odot$ will be shown in the forthcoming paper
(Yoshida et al. 2015, in preparation).

Finally, we discuss the eccentricity of the binary system formed after
the SN that leaves a secondary NS. Due to the mass ejection from the
system, the binary system obtains the eccentricity, $e$, after the SN
explosion. The eccentricity can be evaluated by \citep{bhat91}
\begin{eqnarray}
e=\frac{M_1^i-M_1^f}{M_1^f+M_\mathrm{NS}},
\end{eqnarray}
where $M_1$ is the mass of exploding star before (after) the explosion
indicated as $i$ ($f$) and $M_\mathrm{NS}$ is the mass of the primary
NS. Here, we assumed that the mass ejection occurs quickly and during
the explosion the positions of these stars do not change. By giving
the ejected mass, $M_\mathrm{ej}=M_1^i-M_1^f=0.3 M_\odot$ and
$M_1^f=M_{NS}=1.3M_\odot$, we get $e\approx0.12$, which is compatible
with one of observed binary NSs, J0737-3039, whose current
eccentricity is 0.088 and estimated eccentricity at birth of second
pulsar is 0.11 \citep{pira05b}. The small center of mass velocity of
this system also implies a small ejecta mass and slow pulsar kick
\citep{dall14}.

\section*{Acknowledgements}
We thank A. Wongwathanarat, M. Tanaka and T. Moriya for discussions.
We also appreciate comments on an earlier draft of this paper by
E. M\"uller, T. Piran, T. Tauris, and A. Tutukov.  YS thanks the Max
Planck Institute for Astrophysics for its hospitality. Numerical
computations in this study were in part carried out on XC30 at CfCA in
NAOJ and SR16000 at YITP in Kyoto University.  YS was supported by
Japan Society for the Promotion of Science (JSPS) postdoctoral
fellowships for research abroad.  KT was supported by research
fellowships of JSPS for Young Scientists. This study was also
supported in part by the Grant-in-Aid for Scientific Research
(Nos. 24244028, 26400220 and 26400271), MEXT SPIRE, and JICFuS.


\end{document}